\documentstyle[aps,preprint]{revtex}  
\begin{document}
\draft

\title {A Deterministic Approach to the Protein Design Problem}
\author{Maksim Skorobogatiy, Hong Guo and Martin J. Zuckermann}
\address{
Centre for the Physics of Materials and Department of Physics, \\
McGill University, Rutherford Building, \\
3600, rue Universit\'{e}, Montr\'eal, Qu\'ebec, Canada H3A 2T8.
}

\date{\today}
\maketitle

\baselineskip 16pt
\begin{abstract}

We have considered the problem of protein design based on
a model where the contact energy between amino acid residues is fitted 
phenomenologically using the Miyazawa--Jernigan matrix. 
Due to the simple form of the contact energy function, an analytical 
prescription is found which allows us to design energetically stable 
sequences for fixed amino acid residues compositions and target
structures. The theoretically obtained sequences are compared with 
real proteins and good correspondence is obtained. Finally we discuss 
the effect of discrepancies in the procedure used to fit the contact 
energy on our theoretical predictions.

\end{abstract}

\vspace{0.5in}


\newpage
\baselineskip 16pt
\section{Introduction}

It is well known that natural proteins fold into their native structures
remarkably easily in spite of the enormous number of possible physical
configurations~\cite{creighton}.  For small proteins the native structure
can be determined by the global minimum of the free energy\cite{anfinsen}. 
It has been conjectured\cite{go,shakhnovich1,wolynes} that protein 
sequences are ``optimized'' such that not only is there a stable unique 
structure for the ground state, but the free energy landscape is
funnel-like which leads to efficient folding kinetics.  
A principle of minimal frustration was proposed\cite{bryngelson1} to 
enforce a selection of the interactions between monomers such that as 
few energetic conflicts occur as possible.  
Among other things, considerable theoretical effort has concentrated on
finding proper models for protein folding and investigating various 
sequencings which lead to fast folding kinetics. In this regard, 
statistical analysis has played a very useful role in identifying the 
most relevant factors which determine the process of protein folding.

A statistical mechanical treatment of the protein folding problem
requires a tractable form for the interactions between 
the various amino acid residues.  One approach is to determine
the contact interactions between each pair of amino acid residues
using experimental data. Since there are $20$ different amino 
acid residues, a total of $210$ such interaction parameters is required
for a complete description. This is the approach of 
Miyazawa and Jernigan\cite{mj}. Another simpler 
approach used by many researchers
is to replace the detailed interaction between amino acid residues
by a minimal two parameter model where one parameter
represents the attractive interactions between 
non-polar groups and the second represents
the interactions between the polar and 
non-polar groups. Clearly the simple models based on the second
approach are easier to analyze and they have been successfully used
for qualitative studies of protein folding but they are still 
far from reality. However, even though more involved
models give reasonably good agreement with
experiment, they are unfortunately difficult to analyze theoretically.
From this point of view, it is useful to introduce a compromise between 
minimal and more complete models, such that the resulting model
is detailed enough to capture most 
of the essential characteristics of protein folding while at the same
time being sufficiently simple for a tractable analytic approach.
Indeed it was shown by Grossberg {\it et. al.},that, when studying the
properties of real proteins using an energy interaction
matrix, sufficiently stable ground states can be still obtained
even if there are some errors in the numerical values used for
this interaction matrix~\cite{grosberg}.

In a recent article\cite{li}, Li, Tang and Wingreen (LTW) suggested
a particularly interesting parameterization of a statistical potential 
which was originally derived from known protein structures. By analyzing 
the Miyazawa-Jernigan (MJ) interaction matrix\cite{mj}, they found that the 
entire $20\times 20$ MJ matrix can be fitted very well by a simple form,
\begin{equation}
E^{\theta \sigma}\ =\ q^{\theta}\ + q^{\sigma}\ +\ \beta q^{\theta} 
q^{\sigma}\ \ \ , \label{eij}
\end{equation}
where $E^{\theta \sigma}$ is the contact energy between amino acid  
residues of type $\theta \in (1,..,20)$ and type $\sigma$, 
$q^{\theta}$ is a negative real number which is assigned to 
amino acid  residues of type $\theta$. In their fit to the MJ matrix,
Li {\it et. al.}~\cite{li} found numerical values lying in the range 
[-3.0, 0.0] for the quantities $\{q^{\theta}\}$. The form for the MJ matrix 
given by Eq.~(\ref{eij}) thus has the following physical interpretation. 
The $(q^{\theta}+q^{\sigma})$ term corresponds to solvent exclusion 
which is responsible for the formation of the hydrophilic surface and the 
hydrophobic core of the folded protein whereas the 
$\beta q^{\theta}q^{\sigma}$ term represents segregation which 
is responsible for the differentiation of secondary structures inside of 
the hydrophobic core. This fitting form, while not necessarily unique, 
reveals the intrinsic regularity of the interactions between the various 
amino acid residues and reduces the total of $210$ interaction parameters to 
essentially $20$. Hence this is clearly a useful formal step in the 
theoretical analysis of protein folding. It is also interesting to notice
that the particular form of the contact energy $E^{\theta\sigma}$, being a
combination of linear and quadratic terms, has also been discussed in previous
protein folding literature\cite{obukhov,goldstein,ramanathan}.

In this work we use this form of contact interaction with the 
fitting parameters given by the work of Li, Tang and Wingreen
to examine the following questions {\it analytically}.
For a given protein compact target structure and a 
given amino acid composition, how can we find a sequence of the parameters 
\{$q^{\theta}$\} which minimizes the total energy ? 
Once a sequence which gives the minimum energy for the target
structure is predicted, how does this ``optimal'' sequence compare with 
the protein sequence in the native state ? How sensitive are our 
predictions to the fitted form given by (\ref{eij}) ?  Obviously these are 
important questions related to the protein folding problem.  The 
motivation to investigate these questions analytically comes
from the inspiring work of Shakhnovich and Gutin\cite{shakhnovich2} who 
devised a numerical approach for the design of stable proteins by 
randomly permuting the amino acid residues for a given target structure
using a Monte Carlo algorithm.  We use the 
considerable intuition obtained from various important numerical 
calculations\cite{shakhnovich2,dill1,honeycutt} to serve as
a guide for our {\it analytical} examination of the above questions.

Because we are mainly concerned with the energetics of protein design, 
the sequence will be specified only by the parameters $\{q^{\theta}\}$. Hence 
a given composition is equivalent to a particular set of values for these 
parameters. We next fix a target structure for the protein. This is 
basically a three-dimensional space curve along which nodes labeled by
$i \in (1,..,L)$ are located at equal distances from each other. 
Here $L$ is the total number of monomers (amino acid residues) in the protein
chain. A ``model protein'' is obtained by placing amino acid residues on these 
nodes. Since each amino acid residue corresponds to a specific value of the 
parameter $q^{\theta}$, a sequence of these parameters must be obtained such 
that the total energy is a minimum for the given target structure. This 
sequence can be achieved by permuting the amino acid residues on the nodes 
until the energy reaches a minimum. However such an exhaustive search 
quickly becomes intractable in the limit of long proteins.  Numerically 
one can improve the search using importance sampling techniques
such as the Monte Carlo methods\cite{shakhnovich2}. 
Once the minimum energy configuration is found, we will have obtained,
at least theoretically, a ``model protein'' which is stable energetically.  
Clearly if {\it nature} produces proteins only according to energy
minimization, our ``model protein'' obtained in this manner should be very
similar to the native state of the actual protein. We will thus compare our
predicted amino acid sequence for the given target structure with the native 
state of the corresponding protein as given in the Protein Data Banks (PDB).
Clearly some differences should be expected since proteins also possess 
functional properties and they can not be considered as purely energetic 
units. 

In our analytical work, we use the expression of Eq. (\ref{eij}) as a model 
(referred in the following as LTW model) for the interaction matrix between 
monomers. It is then reasonably straightforward to make some general 
statements concerning the above questions while maintaining a good 
correspondence with the behavior of real proteins. In particular, we derive 
an expression for sequencing the parameters $\{q^{\theta}\}$ for a given target 
structure such that the total energy of the protein is minimized. Our 
expression is exact if the segregation term is neglected. We also show that
the segregation can be included in an extremely good approximation which we
confirm by comparing results of our analytical predictions to those from
exact numerical exhaustive search. Finally, we confirm the results of our 
calculation by numerically calculating the overlap between predicted 
sequences and the native protein sequences using $84$ randomly chosen 
proteins from the Brookhaven Protein Data Bank with lengths ranging 
from $L=21$ to $L=680$. 

The article is organized as follows.  In section II we first derive the 
relevant formula for sequencing without the segregation term and we next 
treat the segregation as a perturbation. In section III we present our
numerical tests on the perturbation treatment of the segregation term.  
The comparison of our predictions to those from PDB will then be presented.
Section IV includes an estimate of the range of validity of our 
predictions in relation to the possibility of discrepancies in the fitted 
form of the MJ matrix as given by Eq.~(\ref{eij}). A summary of the main
results is included in the last section.

\section{Proteins design using the LTW model}

In the following we use the LTW model of Eq. (\ref{eij}) to ``design''
a stable sequence for a ``model protein'' with respect to a given amino acid 
composition and a given target structure. It is worth noting that the 
calculations presented below do not require the presence of a lattice, 
although they can also be applied to lattice models. For our problem,
while we should denote $q_i^{\theta}$ as the parameter $q^{\theta}$ on node
$i$ of the target structure, without causing confusion from now on we shall 
simplify notation by dropping the Greek superscripts. Thus the contact 
interaction between monomer $i$ and monomer $j$ is written as 
\begin{equation}
E_{ij}\ =\ q_i\ + q_j\ +\ \beta q_i q_j\ \ \ , \label{eij1}
\end{equation}
with the understanding that $q_{i}$ is given by one of the twenty possible
values. As mentioned in the original work of Ref. \cite{li}, it follows 
from the values of the fitted $q$-parameters that the solvent exclusion 
term ($q_{i}+q_{j}$) gives the main contribution to the MJ energies $E_{ij}$. 
Hence it is reasonable to consider first the interaction 
$E_{ij}=q_{i}+q_{j}$ only and then investigate the influence of the 
segregation term $\beta q_{i}q_{j}$. This will be our approach.

\subsection{The solvent exclusion term}

As stated above, we first examine the LTW model for the case when the 
interaction matrix between monomers is given by the solvent exclusion term 
$E_{ij}=q_{i}+q_{j}$ only. Here we will assume that two monomers are in contact 
if the distance between them is smaller than a length scale of the order of 
a few angstroms. Following Li {\it et. al.}~\cite{li} we take this scale to 
be $6.5\AA$. Then, if $n_{i}$ denotes the number of closest neighbors to the 
$i$th node on our target structure, the total energy of the protein structure 
is given by
\begin{equation}
E = \sum_{i,j} E_{ij} = \sum_{i,j} q_{i}+q_{j} = \sum_{i} n_{i}q_{i}\\
\label{solvent}
\end{equation}
and $\sum_{i} n_{i} = 2N$ and $N$ is the total number of contacts. 

As an example, we apply Eq.~(\ref{solvent}) to the target structure
with twelve nodes on a $2$D lattice shown in Fig.~(\ref{fig1}). 
By placing twelve amino acid residues (monomers) with parameters 
$q_1,q_2,\cdot\cdot\cdot,q_{12}$ on these nodes, we obtain a 
``model protein''.  For this structure there are 6 pairs of contacts:
the monomer on the first node with ``amino acid residue'' $q_1$ has 
three contacts, monomer on the second node with $q_2$ has two contacts 
while all others have either one or no contacts. 
Using Eq. (\ref{solvent}) the energy is given by 
$E=3q_1+2q_2+q_3+q_4+q_6+q_8+q_9+q_{11}+q_{12}$.
This example shows that it is natural 
to specify the target structure by a vector with elements representing the 
number of closest contacts to each node. Hence for a particular structure with 
$L$ nodes on a $2$D lattice, the geometric conformation is represented by 
the ``contact vector'' $\vec{n}\equiv\{n_i\}$ where $n_{i} \subset \{0,1,2\}$ 
if $i\subset \{2,...,L-1\}$;  and $n_{i} \subset \{0,1,2,3\}$ if 
$i\subset \{1,L\}$. In this notation the $i$th component of $\vec{n}$ gives 
the number of closest neighbors to the $i$th node. A similar prescription can 
easily be written down for $3$D systems. It is clear that the {\it length} 
of a vector $\vec{n}$ will in general increase as the number of contacts in a 
given structure increases.

Next we introduce a second vector with $L$ components, $\vec{q}\equiv \{q_i\}$, 
which specifies a particular sequence of the values of $\{q_i\}$ imposed on 
the geometrical structure defined by $\vec{n}$. The placing of amino acid 
residues on the nodes of the target structure is equivalent to assigning the 
corresponding values of $q_i$ to each node. Then using Eq. (\ref{solvent}) 
the energy of the ``model protein'' can be rewritten as
\begin{equation}
E = \sum_{i} n_{i}q_{i} = \vec{n}\cdot \vec{q}\ \ \ .
\label{solvent1}
\end{equation}
Eq.~(\ref{solvent1}) shows that the energy is separable in the geometrical 
factors and the details of a particular sequence in this model. Using our 
notation, if one draws all possible vectors of type $\vec{n}$ corresponding 
to all different geometrical structures of a protein, the configurational 
space of the protein will be represented by a vector bundle generated by 
the set of all $\vec{n}$ vectors, while a particular sequence will be 
represented by a single vector $\vec{q}$. Then, as seen from 
Eq.~(\ref{solvent1}), the energy spectrum for a particular set of amino 
acid residues on a given target structure will be determined by the 
projection of the vector bundle onto the vector $\vec{q}$. As mentioned above, 
more compact conformations will in general have longer $\vec{n}$ vectors 
since compact structures tend to have more contacts between monomers and the 
corresponding length will be proportional to $L^{\frac{1}{2}}$. Given that 
the most compact conformation is represented by the longest vector, 
$\vec{n}_{max}$, all less compact conformations will lie inside a sphere of 
radius equal to the magnitude of $\vec{n}_{max}$. Nearly compact conformations 
will then lie in the neighborhood of this sphere\cite{foot1}. This is shown 
in Fig.~(\ref{fig2}).

We now consider a specific sequence defined by a vector $\vec{q}$ with 
magnitude $Q$. We next denote a given compact geometrical target structure 
for a given protein as $\vec{n}_{\alpha}$. If we are {\it not} limited by a 
particular composition represented by a fixed set of values of the parameters 
$\{q_i\}$, the energy minimization and design is straightforward.  
Eq.~(\ref{solvent1}) shows directly that the system energy is minimized if we 
choose $\vec{q}$ to be anti-parallel to the vector $\vec{n}_{\alpha}$,
as shown in Fig. (\ref{fig2}). For this trivial case we thus obtain
the ``ideal'' sequence, $\vec{q}_{ideal}$, which gives the lowest possible 
energy for the target structure represented by $\vec{n}_{\alpha}$
\begin{equation}
\vec{q}_{ideal}=\vec{n}_{\alpha}(\frac{Q^{2}}{\vec{n}_{\alpha}\cdot
\vec{n}_{\alpha}})^{\frac{1}{2}} \ \ .
\label{qideal}
\end{equation}

However such an  ``ideal model protein'' is not realistic as it does not 
respect the particular values of the parameters $\{q_i\}$ corresponding to 
the actual set of amino acid residues defining the primary structure of the 
protein. Hence the set of values of the parameters $\{q_i\}$ must be fixed 
in order to specify a particular protein. The problem then becomes more 
complicated as the total energy must now be minimized subject to this constraint. 
Furthermore we can only minimize the energy by shuffling the given parameter
set of the parameters \{$q_i$\} among the different nodes of the target 
structure. This corresponds to performing discrete rotations of the vector 
$\vec{q}$ in its specific vector space rather than the continuous rotations 
used to find the ``ideal'' sequence. However, even under this constraint we 
can still solve the problem close to analytically.  

We begin by stating the following well known inequality. If 
\begin{eqnarray*} 
n_{1}\geq n_{2}\geq n_{3}...\geq n_{L}\\
q_{1}\geq q_{2}\geq q_{3}...\geq q_{L}
\end{eqnarray*}
where $n_{i}$ and $q_{i}$ are arbitrary real numbers, then 
\begin{equation}
\sum^{L}_{i=1} n_{i}q_{i}\geq \sum^{L}_{i=1} n_{i}q_{k_{i}}
\label{inequ}
\end{equation}
where $q_{k_{i}}$ represents any permutation of the set of parameters $q_{i}$.
We now fix the amino-acid composition by fixing the values of the components, 
$q_i$, of a given vector $\vec{q}$, where the amino-acid residue corresponding 
to $q_i$ is placed on the $i$th node of the target structure. We change the 
amino-acid sequence on the target structure by permuting the values $q_i$ 
among the nodes of the target structure represented by the vector 
$\vec{n}_{\alpha}$. This gives us a new vector representing a different 
amino-acid sequence on the target structure. Now the minimization of the 
energy with respect to the target structure given by 
$E=\vec{n}_{\alpha}\cdot\vec{q}$ clearly requires us to find a vector,
$\vec{q}_{min}$, which is as anti-parallel to $\vec{n}_{\alpha}$ as possible. 
Since the fitted numerical values for the parameters $q_i$ are all negative, 
we devise the following procedure on the basis of the inequality of 
Eq.~(\ref{inequ}). In this procedure we minimize the energy by first sorting 
all the components $q_{i}$ according to their absolute values, and then 
sorting all nodes on the geometrical target structure according their number 
of nearest neighbors. We then place the amino acid residues corresponding to 
larger values of $|q_i|$ on the nodes of the target structure which have 
larger numbers of contacts, and the amino acid residues corresponding to 
smaller values of $|q_i|$ on the nodes with a smaller number of contacts. 
This is a systematic way of finding the sequence which gives a stable 
(minimum energy) protein target structure. Since no exhaustive search is 
involved, essentially no computer is needed. However, from the energy point 
of view this procedure, because it is based on Eq. (\ref{solvent}), 
will produce sequences of $\{q_i\}$ in which as many hydrophilic 
amino acid residues as possible are on the surface of the target structure 
and as many hydrophobic amino acid residues as possible are in its interior. 
This will be corrected in the next section when the segregation term is 
included.

The method discussed here may lead to degeneracies of the final sequence 
$\{q_i\}$, {\it i.e.} there may be more than one sequence which gives the 
same minimum energy for the target structure $\vec{n}_{\alpha}$.
This is easy to understand by noticing that a compact structure 
predominantly consists of ``surface'' monomers with a small number of 
nearest neighbors and ``interior'' monomers with a large number of such 
neighbors. Thus the geometrical permutation of ``interior'' monomers among 
themselves or ``surface'' monomers among themselves made by permuting the 
relevant parameters, $q_i$, will not alone change the energy. The sequence 
obtained from the above procedure hence specifies the structure up to a 
differentiation of ``surface'' and ``interior'' monomers only.

\subsection{The segregation term}

In order to treat the segregation term $\beta q_{i}q_{j}$ in our analysis, we 
use the result of Ref.~\cite{li} that it is small in comparison with  
the solvent exclusion term studied in the last subsection and that it should 
not alter the relationship between ``surface'' and ``interior'' monomers. 
Thus the inclusion of this term should not cause a substantial change in the 
geometric vector, $\vec n$, of our model protein. However this term will at 
least partially break the degeneracy in the secondary structure. Because the 
segregation term is quadratic in $\vec{q}$, it energetically favors segregation 
between different amino acid residues and leads to an increased specificity 
of the overall structure. Including this term, we now investigate which 
sequence should be chosen so that the given target structure will have 
minimal energy. 

As in the previous section we first find the ideal sequence $\vec{q}_{ideal}$ 
by only fixing its length to be $Q$ and then minimizing the energy of the 
target structure denoted by vector $\vec{n}_{\alpha}$. 
Mathematically the problem is to minimize the function 
$\sum_{ij} E_{ij}=\sum_{ij} q_{i}+q_{j}+\beta  q_{i}q_{j}$, $\beta < 0$, 
$q_{i} < 0$, while keeping the length of $\vec{q}$ fixed at the value $Q$.
For this purpose, we introduce the contact matrix, $C^{\alpha}$, which has 
elements $C_{ij}^{\alpha}=0$ if $i$-th monomer does not have $j$-th monomer 
as a nearest neighbor and $C_{ij}^{\alpha}=1$ otherwise. We can then rewrite 
the energy of the target structure with a sequence $\vec{q}$ as follows 
\begin{equation}
\left \{ \begin{array}{lll}
E^{(\alpha)} & = & \vec{n}_{\alpha} \vec{q}+\frac{\beta}{2} \vec{q} 
C^{\alpha} \vec{q}\\ \vec{q}\cdot \vec{q} & = & Q^{2}\\ 
\end{array}
\right .
\label{eq5}
\end{equation}
Using the method of Lagrange multipliers one finds the following sequence 
which gives the minimum energy, 
\begin{equation}
\vec{q}_{ideal}=[\lambda I-\beta C^{\alpha}]^{-1}\cdot \vec{n}_{\alpha}
\label{qideal1}
\end{equation}
where $I$ is the identity matrix and $\lambda$ is the solution of equation
\begin{equation}
Q^{2}=\vec{n}_{\alpha}\cdot [\lambda I-\beta C^{\alpha}]^{-2}\cdot 
\vec{n}_{\alpha}\ \ . \end{equation}
These equations can easily be solved and they can be replaced by their 
expansions in $\beta$ for proteins with small number of monomers. 

The sequence $\vec{q}_{ideal}$ as solved above gives the lowest possible
energy for a given target structure. However it does not respect the 
actual amino acid composition of the real protein. We should instead fix 
the composition and only shuffle the elements, $q_i$, of the vector $\vec{q}$ 
instead of changing their values. In the previous subsection, we gave a 
prescription for finding a sequence {\it exactly} for a given composition. 
However with the inclusion of the segregation term we can no longer access 
an exact solution, but we can make an extremely good approximation for the 
desired sequence. The approximation we use here has the character of
a mean field analysis in that we replace one of the vectors, $\vec{q}$, in 
the quadratic term of Eq. (\ref{eq5}) by $\vec{q}_{mf}$, {\it i.e.}
\begin{equation}
E^{(\alpha)}\ \approx\ (\vec{n}_{\alpha} + \frac{\beta}{2} 
\vec{q}_{mf} C^{\alpha} )\cdot 
\vec{q} \ \equiv\ \vec{n}_{\alpha}^{mf}\cdot \vec{q}\ \ .
\label{eq8}
\end{equation}
Here $\vec{n}_{\alpha}^{mf}$ is the sum inside the bracket and results in a 
slight change of the elements of the nearest neighbor vector. The choice of 
$\vec{q}_{mf}$ may be $\vec{q}_{ideal}$ using Eq. (\ref{qideal1}) 
which we obtained in this section; or the {\it optimized} sequence which 
we obtained without the segregation term. With this new form of energy, we 
can use the same sorting prescription discussed at the end of the last 
subsection to find the minimal sequence $\vec{q}$ under the constraint of 
fixed composition. In the next section we shall confirm this mean filed like
analysis by comparing results to those from the exhaustive numerical search.

\section{Numerical results}

To confirm the predictions by our analytical design method discussed in the
last section, in the following we shall examine the quality of our mean field
like analysis of the segregation term and compare our results to those from
real protein structures using $84$ randomly chosen proteins from PDB.  

\subsection{Lattice enumeration}

While our method of finding an optimal sequence without segregation term
was exact, the validity of the mean field approach to include the
segregation, Eq. (\ref{eq8}), needs to be investigated. 
To this purpose, we have used the compact structure of Fig.~(\ref{fig1}) as the 
target structure $\vec{n}_{\alpha}$, and fixing a composition of the 
parameters $q_i \in (-4.0,0.0)$ we designed the optimal sequence 
$\vec{q}_{opt}$ using our analytical method including the segregation term.
With a choice of $\vec{q}_{mf}$ (see below), this procedure minimized the 
energy $E^{(\alpha )}\approx \vec{n}_{\alpha}^{mf}\cdot \vec{q}_{opt}$ within 
the mean field approximation to give $\vec{q}_{opt}$ by Eq. (\ref{eq8}).
We then exhaustively generated all other possible structures of this $12$ 
monomer self-avoiding chain on a 2D lattice, and we denote these
structures by $\vec{n}_{\eta}$ where $\eta \neq\alpha$. The energies,
$E^{(\eta )}$, of these other structures are calculated using 
Eq. (\ref{eq5}). We then compare $E^{(\alpha )}$ with the smallest 
$E^{(\eta)}$ to see which is lower. Finally, we have checked $60$ different 
amino-acid compositions with $q_i$'s randomly generated from the above range. 
We have fixed the parameter $\beta=-0.476$ in this numerical check, 
and several observations are in order.

First, for $57$ out of the $60$ random compositions tested, 
$E^{(\alpha )}< E^{(\eta)}$. Hence our analytical design with the mean 
field approximation indeed generated the sequence which guarantees the 
compact target structure to be the ground state. For the other $3$ 
compositions of the $q_i$'s, each case has only one chain structure which 
gave slightly lower energy than $E^{(\alpha )}$. However the difference 
is very small, being $0.3$\%, $0.6$\%, and $2.8$\%. We may thus conclude 
that the accuracy of our mean field treatment of the segregation term is 
acceptable. Second, we found no difference to this conclusion using two 
different $\vec{q}_{mf}$ in Eq. (\ref{eq8}). The two most evident choices 
of $\vec{q}_{mf}$ are the optimized sequence which 
we obtained by neglecting the segregation term (see section II.A), 
or the $\vec{q}_{ideal}$ of Eq. (\ref{qideal1}) where we included the
segregation term. They work equally well. Finally,
our numerical test also gave a measure of the relative importance of the
segregation term. If we directly use the optimized sequence determined
without the segregation term as $\vec{q}_{opt}$ to compute the energy using
Eq. (\ref{eq5}), rather than determine $\vec{q}_{opt}$ as we have done so far,
the comparison with the lattice enumeration is worse. In this case $27$ 
out of the $60$ random compositions gave $E^{(\alpha )}< E^{(\eta)}$.
On the other hand, the other $33$ compositions produced a chain structure 
with lower energy than the target structure $\vec{n}_{\alpha}$, with 
differences which are less than $10$\%. This means that choosing a sequence 
which is only optimal without the segregation term, we have a lower 
probability of making a target structure to be the ground state, 
although the result is still not far from it.

Our mean field approach can also be applied to the model where the 
contact energy is determined solely by the segregation term:
$E_{i,j} = \beta q_iq_j$. In terms of the total energy, we have 
$E^{(\alpha)}=\frac{\beta}{2} \vec{q} C^{\alpha} \vec{q}$. 
With the approach discussed above, we again tested $60$ randomly generated 
sequence compositions by comparing our analytically designed results to 
the exact enumerations using the $12$ monomer chain of Fig. (\ref{fig1}) 
as the target. Of these, in $49$ cases the designed sequences made the 
energy of the target structure to be the ground state. In the rest $11$ 
cases, the designed 
sequences produced higher energies than some structures other than the 
target, but the difference were less than 10\% with only one case about 
28\%. Hence for this purely quadratic model, our design method also works 
quite well.

With the above comparisons, we may conclude that the mean field approach 
to the segregation term is an acceptable approximation for finding an 
optimal sequence for general models described by contact energies
in the form $E_{i,j}=C_0 + C_1(q_i+q_j) + C_2(q_i q_j)$.

\subsection{Comparison to PDB}

In the previous sections we have derived analytical expressions and 
devised exact or approximate methods to find sequences with fixed 
composition which minimize the total energy of a given target structure. 
Using this method we showed how ``model proteins'' can be generated. 
The next step is to see how closely these ``model proteins" resemble 
the native states of the real proteins with the same composition. This 
is important since natural proteins have other functional properties and 
do not just minimize their energy during the folding process. In addition our 
analysis so far is based on the model described by Eq. (\ref{eij}) which is a
result of fitting to experimental data\cite{li}. Hence some numerical
discrepancies in the fitted parameters, \{$q_i$\}, can be expected.
These will give rise to differences between our model proteins and 
their real counterparts.

In order to check the accuracy of our predicted sequences with fixed 
composition and fixed target structure, we randomly chose $84$ 
proteins with lengths between $L=21$ to $L=680$ from the 
Brookhaven Protein Data Bank as our target structures.  We examine
the quality of the predictions by using two parameters which give a measure
of the overlap of our ``model proteins'' with real protein structures.  
The first parameter uses a scale which only distiguishes between 
hydrophobic and hydrophilic amino acid residues
\begin{equation}
PH_{o}\equiv 1-\frac{1}{L}\sum_{i} |\alpha_{i}-\alpha_{i}^{real}|\ \ .
\label{ph0}
\end{equation}
Here $\alpha_i$ and $\alpha_i^{real}$ equal $1$ if $i$-th amino acid 
residue is hydrophobic and $0$ otherwise. In the following we consider
an amino acid residue to be hydrophobic\cite{li} if its strength 
$q_i \leq -1.5$. Clearly $PH_o$ approaches unity if the predicted 
values of $\alpha_i$ are close to those of the real protein, 
$\alpha_i^{real}$. The second parameter we use is defined as
\begin{equation}
S_o\equiv 1-\left(
\frac{\sum_i |q_i-q_i^{real}|^2}{\sum_i {q_i^{real}}^2}\right)^{\frac{1}{2}}
\ \ ,
\label{s0}
\end{equation}
where $\{q_i\}$ is the predicted sequence while $\{q_i^{real}\}$ is the 
real sequence. This quantity is a more refined measure than $PH_o$ since it
uses the complete $20$ letter code instead of the two letter code used to
define $PH_o$. Throughout the calculations we have used the values 
of $\{q_i\}$ as fitted in Ref. \cite{li}. Again, we consider two monomers 
in contact with each other if the distance between them is less than $6.5\AA$.

Before proceeding any further we present an expression for $PH_o$ for a 
{\it random} sequence of the same composition as a real protein. This 
expression can easily be obtained from Eq.~(\ref{ph0}). If $n_0$ is 
the total number of hydrophilic amino acid residues and $n_1$ is the number 
of hydrophobic ones, we obtain
\begin{equation}
PH_o\ =\ \frac{n_{1}^{2}+n_{0}^{2}}{(n_{1}+n_{0})^{2}}\ \ .
\label{ph01}
\end{equation}
From this expression we conclude that $PH_o$ is usually greater than 
$0.5$ for a random sequence and it equals to $0.5$ when $n_1=n_o$.

The first set of results from our calculations gives the correspondence 
between the exact solution of the model with an interaction matrix given by 
Eq. (\ref{solvent}) where the segregation term is neglected and the
real protein sequences. Using our method described in the 
last section we minimized the energy of geometrical target structures 
taken from PDB while keeping the composition fixed and identical
to that of real proteins. The results for 12 typical proteins are tabulated 
in Table 1. The first column gives values for the $12$ proteins 
randomly selected from PDB. The second column gives values for the two 
letter-code measure, $PH_o$, as obtained by using our minimization procedure.
The data shows that the ``model protein'' sequences thus obtained
have a 61\% to 71\% correspondence with real proteins for the two 
letter-code measure. On the other hand the best random sequence
(third column) gives only a 59\% correspondence. Since 
the model sequences are degenerate as discussed before, we computed all
possible degenerate sequences and the corresponding $PH_o$ 
values\cite{foot2}. The best and the worst correspondences with real 
proteins are listed in column four.  From these numerical values we 
conclude that, when using the two letter-code measure, $PH_o$, 
even the simple model with only the solvent exclusion term can lead 
to good correspondence between our predicted ``model protein'' and the 
real protein.  On the other hand, if we use the more stringent measure 
$S_o$ which is based on the twenty letter-code, the ``model protein'' 
and the real protein have considerably less overlap as shown in the fifth 
column of Table I.  Here the best case is less than 50\%.  

The overlap is greatly improved if the segregation term is included.
This term gives an overlap parameter $S_o$ of $0.87$ between 
$\vec{q}_{ideal}(\beta=0)$ and $\vec{q}_{ideal}(\beta=-0.467)$. This 
implies that the inclusion of the segregation term will give similar 
energy behavior and we therefore do not expect any significant changes 
when using the solvent exclusion term only. On the other hand, inclusion 
of the segregation term should improve the $S_o$ overlap parameter 
with real protein sequences because this term lifts the degeneracies 
in the position of amino acid residues. Using our analytical design procedure 
we obtained the results listed in Table II. In particular, in comparison with
Table I, the parameters $PH_o$ do not change very much even though the 
degeneracies discussed in Ref. \cite{foot2} have largely been lifted.
When the segregation term is included, the difference between the best and 
worst limits of Table I decreased to around 1\% only due to the lifting of 
the degeneracies. On the other hand, column four of Table II shows clearly
that the twenty letter-code measure $S_o$ has been significantly improved 
by about a factor of two to values ranging from $0.36$ to $0.53$ when the 
segregation term is present.

The last three columns of Table II give the total energies of the twelve 
proteins.  Column five gives the energies for the real protein structures 
as computed using the MJ matrix. Columns six and seven correspond to the 
predicted ``model protein'' structure when the MJ matrix or equivalently the 
LTW model Eq.~(\ref{eij}) is used to calculate the  energies. This column 
shows that theoretical predictions using the MJ interaction matrix 
give lower energies than those of the real proteins by 20\%-42\% 
for almost all the proteins tested. The exception is the protein coded 
{\it 1ed} for which the predicted energy is higher than that of the true 
native state by 5\%. Similar behavior was found using the LTW model, 
but in this case there are three proteins for which the theory prediction 
gives higher energies than for the native states.
In this regard the difference between the two theory predictions reflects
the quality of the fit of the MJ matrix as given in Eq.~(\ref{eij}). 

The complete data for the energy comparisons of $84$ proteins with 
their ``model" counterparts is presented in Fig.(\ref{fig3}). 
The horizontal axis corresponds to the number of monomers in a 
chosen protein. The vertical axis gives the percentage difference in 
energies between real proteins and the corresponding ``model" or 
``designed" proteins. Several comments concerning this figure need to be 
made. First of all, the proteins examined for this figure were chosen 
randomly from the Brookhaven Protein Data Bank and we concentrated 
on those proteins consisting of one chain and representing autonomous units. 
An interesting feature of Fig.(\ref{fig3}) is the broad scatter of points for 
short proteins on an energy difference scale along the vertical axis. 
From general arguments it is clear that the energy optimization of a 
structure should be much easier in nature for shorter proteins than for 
longer proteins. This is why we expect a greater similarity between
real and ``model'' proteins for low monomer number. This tendency
is examined in Fig.(\ref{fig3}) which shows that not all of the short 
proteins chosen have good correspondence with the related ``model"
proteins.  A possible explanation for the broad scatter
is that the energy of a short protein is much more sensitive 
to structural details such as side chains, rigid bonds etc. 
than the energy of a long protein. Such structural details are not 
considered in simple theories and they clearly
impose additional constraints on the geometrical
and energy landscapes of a protein. For the number of 
interactions present in short proteins these 
structural details can be expected to give substantial deviations 
from the ``model proteins'' which were obtained
by energy minimization alone. In contrast the large 
number of interactions present in long proteins 
can be expected to suppress the influence of structural details on their 
energies.  This implies that there should be less scatter in the energy
difference between real proteins and ``model proteins" obtained from energy
minimization using simple theories in the case of high monomer number.

In order to clarify these considerations we investigated the structure of 
two best case scenarios (1edn, 1hpt) and two worst case scenarios 
(1r69, 2cym) for short proteins. These structures are shown in Fig. 
(\ref{fig4}).  We notice that while 1edn and 1hpt are very compact
proteins, 1r69 and 2cym have some interior cavities related to their
function (1r69 for example is an amino-terminal domain). 
The presence of structural cavities is probably due to packing arrangements 
among the amino acid residues. This packing constraint plays a role of 
a ``structural perturbation" from the ``unperturbed" state defined by 
the minimal energy requirements of abstract point residues interacting 
via MJ matrix as used in our simple model. In contrast,  
for the 1edn protein, the abundance of hydrophilic groups leads to 
a very compact, energetically minimized structure where 
there are no ``structural perturbations".           
           
Even though there can be other contributions to the energy differences
between ``model" and real proteins,
the energy comparison with the true native structure of real proteins 
clearly suggests that while energy minimization plays an important role in 
protein folding, it is definitely not the only rule that nature follows.

\section{Discussion}

We have shown that LTW model and the related analytic procedure for the 
design of stable ``model proteins'' leads to reasonable results which  
compare well with real proteins for the two measures used.

The one point of concern in this discussion relates to the actual fit of the 
MJ matrix by Li. {\it et. al.}\cite{li} which was based on Eq. (\ref{eij})
and which resulted in the specific values of the parameters $\{q_i\}$ used 
here in the analysis for protein design. However we recognize that
there is always some small uncertainties to any numerical fit, hence it is
useful to determine as to what extent these small uncertainties affect
on the predictions and conclusions of sections II and III. 
To this purpose, we notice that using quantities defined as 
$\delta q_{i}^{j}\equiv \frac{MJ(i,j)-E_{i,j}}{1+\beta q_{j}}$,
then substitution of $q_{i}+\delta q_{i}^{j}$ into Eq. (\ref{eij}) gives the 
correct values of the elements of the MJ energy matrix, \{$MJ(i,j)$\}. 
This suggests that we may use 
$\delta q_{i}=\frac{\sum_{j=1}^{20} |\delta q_{i}^{j}|}{20}$ as a measure
of the fitting quality of the parameter $q_i$. Obviously,
the better the fit, the smaller the value of  $\frac{\delta q_{i}}{|q{i}|}$.
For the LTW fit, this quantity is not greater than 20\%. Now let's
assume that there are $L$ monomers in a compact target structure,
and that the errors for each parameter $q_{i}$ are independent.  
For a compact conformation there are $\sim dL^\frac{d-1}{d}$ 
``surface'' monomers and there are $\sim L$ ``interior'' monomers. 
When the monomers are shuffled in order to find the most
stable sequence as discussed in section II, the possible difference 
in energy given by the LTW model will be of the order of 
$\Delta E\sim dL^\frac{d-1}{d}<q>n_{s}$ where $<q>$ is the average of the 
parameters $\{q_i\}$ and $n_{s}$ is close to the average number of 
nearest neighbors for the ``surface'' monomers.
In the fit of Ref.~\cite{li}, $<q>$ is of the order 
of unity. On the other hand, the errors in the energy due to the small
discrepancies $\delta q_i$ are of the order of 
$\delta E\sim\delta(\vec{n}\cdot\vec{q})\sim L\frac{<\delta 
q>n_{b}}{20^{1\over 2}}$ since there are 20 independent parameters $q_{i}$. 
Here $<\delta q>$ is the average of the quantities $\delta q_i$.  
Clearly if $\delta E\sim \Delta E$, we cannot make any reasonable predictions. 
From this discussion we conclude that the use of the LTW model for our 
calculation is justified if 
\begin{equation}
L \ll (20^{1\over 2}\frac{n_{s}}{n_{b}}\ \ d\frac{<q>}{<\delta q>})^{d}\ \ 
\ \ . \label{condition1}
\end{equation}
Hence, as the fitting uncertainty is at most 20\%, {\it i.e.}
$\frac{<q>}{<\delta q>}\sim 5$, our predictions should be valid for
$L \ll 500$ in 2D; and $L \ll 5000$ in 3D. Thus even for a 20\% error 
in the parameters $\{q_i\}$, our procedure based on the LTW model can 
still describe and make predictions for relatively long proteins.

\section{Summary}

In this work we applied the model of Li, Tang, and Wingreen\cite{li} 
to design ``model proteins'' which have minimum energy for a fixed 
amino acid composition and a given target structure. The model is
well suited to this procedure because it is reasonably 
accurate and yet sufficiently simple for analytical or deterministic
calculations. Using the vector notation of section II for 
the target structure and the sequence, we were able to find a simple 
method which determines model protein sequences based on the LTW 
model. We estimated that our method can be applied to protein 
chains with a few hundred monomers even if there are substantial errors 
in the parameters $\{q_i\}$.  Using $84$ randomly chosen real proteins 
from protein data banks we confirmed that our predicted sequences are 
reasonably realistic. Furthermore our ``model protein'' sequences 
have total energies as computed from the LTW model or from the
original MJ matrix which are for most cases lower than those of the real 
proteins.  While several factors could be responsible for this difference, 
it suggests that energy minimization is indeed important but it is not
the only factor which determines the native structure of proteins.
We have also found that the segregation term in the LTW model plays an
important role of lifting the degeneracies of the sequence which occur
when only the solvent exclusion term is included in our calculations. 
In addition this term improves the $20$ letter-code comparison between 
our theory and real proteins by a major factor.  On the other hand 
for the two letter-code overlap measure used in section III to compare 
with real proteins, our predictions for the proteins examined here
are usually 10\% (the best 19\%) better than those obtained by 
using a random sequence. As pointed out in Ref. \cite{shakhnovich2}, 
one does not expect a 100\% homology between the real and the predicted 
sequence for the design problem as degeneracies and ``structural 
perturbations'' are present. The merit of our method lies in the fact 
that it is easy to use, allows analytical or partially analytical solutions 
of the problem, gives simple geometrical interpretation of the results, 
and uses essentially no computer time while giving reasonable comparisons 
with real proteins.  

\section*{Acknowledgments}
We gratefully acknowledge support by the Natural Sciences and
Engineering Research Council of Canada and le Fonds pour la Formation de
Chercheurs et l'Aide \`a la Recherche de la Province du Qu\'ebec.

\begin{figure}
\caption{
Sketch of a typical target structure on a 2D lattice with $12$ 
nodes. The notation $n$($q_m$) on each node states that the node 
associated with parameter $q_m$ has $n$ nearest neighbors.
\label{fig1}}
\end{figure}

\begin{figure}
\caption{
The configurational space of the protein as represented by a 
vector bundle generated from all vectors of type $\vec{n}$. A 
particular sequence is represented by a vector $\vec{q}$. 
\label{fig2}}
\end{figure}

\begin{figure}
\caption{
The percentage energy difference between real proteins and ``model"
proteins versus the number of monomers in a protein. A value of
zero on the scale of the ordinate is equivalent to $100$\% correspondence. 
\label{fig3}}
\end{figure}

\begin{figure}
\caption{
The following native protein conformations were taken
from the PDB: (a)~1edn, (b)~1hpt, (c)~1r69, (d)~2cym.  
\label{fig4}}
\end{figure}


\begin{thebibliography}{99}
\bibitem{foot1}
The length of the most compact conformations can slightly vary 
due to the differences in the distribution between monomers in the 
``interior'' and those on the ``surface'' of the compact structure, but 
we can neglect this variation when the protein length $L$ is large enough,
\bibitem{foot2}
The origin of degeneracy in the absence of the segregation term
can easily be explained.  In this case the 
energy of a structure $\vec{n}$ with sequence $\vec{q}$ is    
$E=\vec{n}\cdot \vec{q}$. The $i$-th component of $\vec{n}$ 
corresponds to the number of nearest neighbors of the $i$-th monomer.
Usually, $n_{i}$ varies in the range of $0$ to $12$ taking only 
integer values.  A permutation of amino acid residues 
corresponding to monomers with the same number of nearest neighbors 
will clearly not lead to a change in energy since the solvent exclusion term
is additive, which leads to the degeneracy with respect to the 
shuffling of amino acid residues.  Although the energy remains unchanged,
shuffling these amino acid residues will influence the overlap parameter 
$PH_o$.
\end{thebibliography}
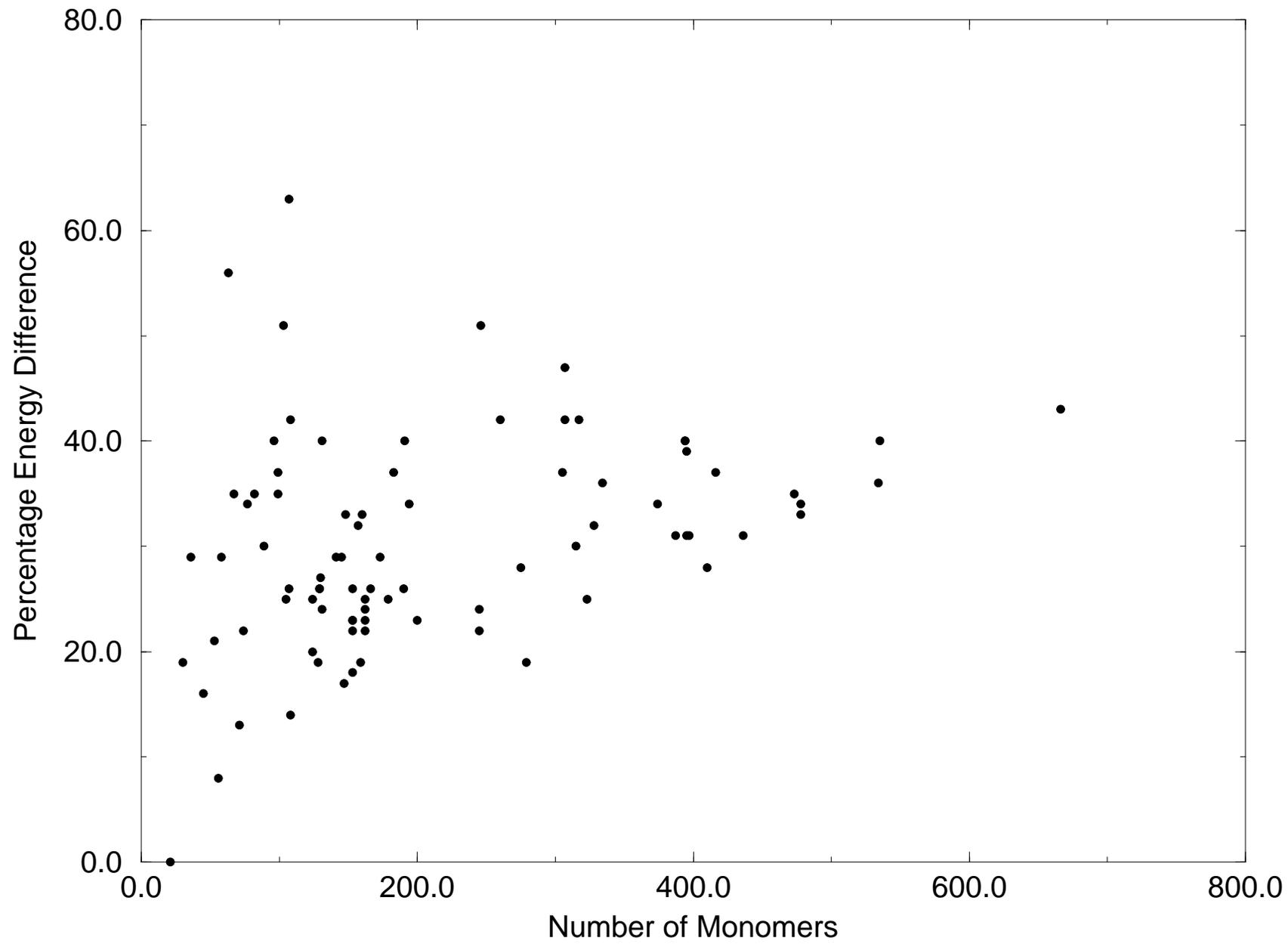
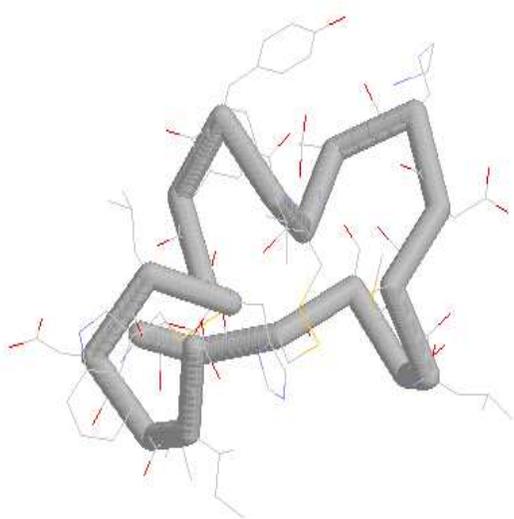
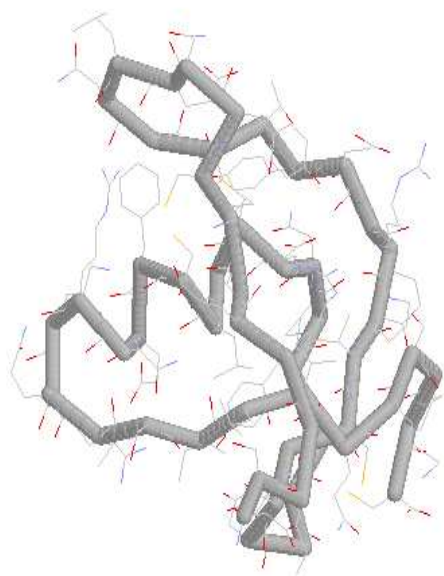
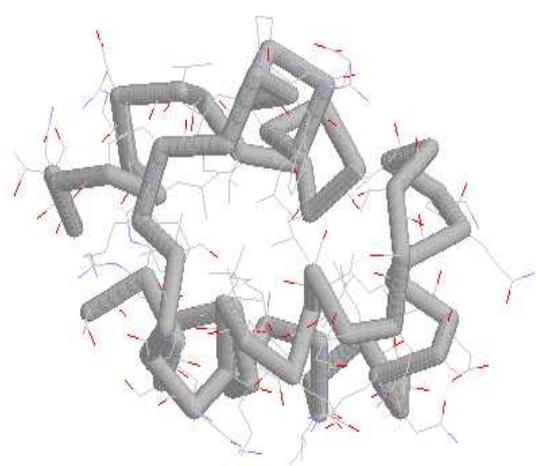
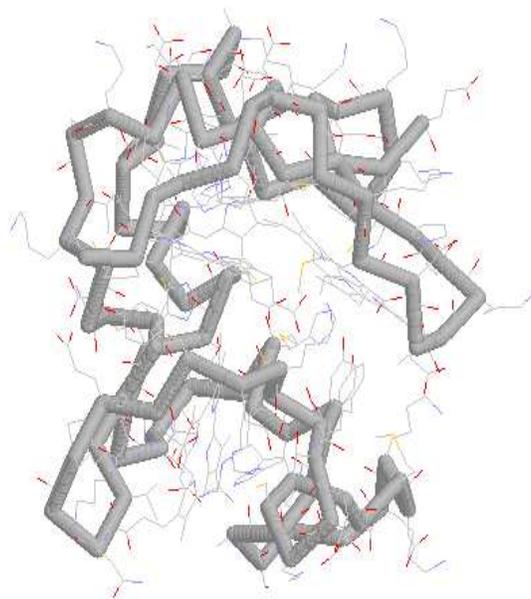
\end{document}

\renewcommand{\thetable}{\Roman{table}}
%, here is table 1
%  A sample of table 
%
\begin{table}
\caption{Values of the overlap between real protein structures and those
of the predictions. Only the solvent exclusion term is included in the
analysis. Column 1: protein codes from PDB;  column 2: the measure $PH_o$
obtained from the analytical predictions; column 3: $PH_o$ from random
sequences; column 4: the best and worst predicted $PH_o$ values among all
the degenerate sequences; column 5: the twenty-letter code overlap $S_o$.
}
\vspace{0.1in}
\begin{center} 
\fbox{
\begin{tabular}{ccccc}
Protein& $PH_o$   & $PH_o$ random & $PH_o$ minimized& $S_o$\\ 
code   & minimized& sequence      & degeneracy limits& \\ \hline\hline
621p  & 65\% & 54\% & 71\%$\geq$ 56\% & 28\% \\
\\
129l  & 66\% & 53\% & 73\%$\geq$ 59\% & 27\% \\                 
\\
4mbn  & 62\% & 57\% &    62\%         & 23\% \\                 
\\
144l  & 64\% & 55\% & 72\%$\geq$ 58\% & 26\% \\                 
\\
451c  & 61\% & 59\% & 61\%$\geq$ 59\% & 20\% \\                 
\\
181l  & 64\% & 56\% & 70\%$\geq$ 62\% & 26\% \\                 
\\
7api  & 64\% & 55\% & 67\%$\geq$ 59\% & 25\% \\                 
\\
6dfr  & 70\% & 55\% & 80\%$\geq$ 61\% & 33\% \\                 
\\
2pal  & 67\% & 58\% & 70\%$\geq$ 64\% & 21\% \\                 
\\
2gch  & 63\% & 56\% & 67\%$\geq$ 61\% & 28\% \\                 
\\
1edn  & 71\% & 53\% & 81\%$\geq$ 71\% & 47\% \\                 
\\
1epg  & 66\% & 51\% & 70\%$\geq$ 58\% & 37\% \\                 
\\

\end{tabular}}
\end{center}
\end{table}
 
\newpage
\begin{table}
\caption{Values of the overlap between real protein structures and those
of the predictions. Both the solvent exclusion and segregation terms
are included in the analysis. Column 1: protein codes from PDB;  
column 2: the measure $PH_o$ obtained from the analytical predictions; 
column 3: $PH_o$ from random sequences; column 4: the twenty-letter code 
overlap $S_o$;  column 5: energy of the real protein sequence as computed 
using the MJ matrix;  column 6: energy of the predicted sequence computed
from the MJ matrix;  column 7: energy of the predicted sequence computed
from the LTW fitting form.}
\vspace{0.1in}
\begin{center} 
\fbox{
\begin{tabular}{ccccccc}
Protein&$PH_o$&$PH_o$ &$S_o$&Energy &Energy &Energy \\ 
code& minimized& random& &real protein &using MJ &using LTW\\
\hline\hline
621p  & 65\% & 54\% & 42\% & -1123 & -1414 & -1326\\
\\
129l  & 67\% & 53\% & 41\% & -1112 & -1378 & -1622\\                 
\\
4mbn  & 62\% & 57\% & 34\% & -1131 & -1430 & -1322\\                 
\\
144l  & 66\% & 55\% & 40\% & -1101 & -1353 & -1648\\                 
\\
451c  & 56\% & 59\% & 35\% & -395  & -563 & -465\\                 
\\
181l  & 64\% & 56\% & 39\% & -1110 & -1393 & -1781\\                 
\\
7api  & 63\% & 55\% & 36\% & -2572 & -3376 & -3037\\                 
\\
6dfr  & 66\% & 55\% & 42\% & -993 & -1184 & -1174\\                 
\\
2pal  & 67\% & 58\% & 35\% & -563 & -801 & -529\\                 
\\
2gch  & 63\% & 56\% & 41\% & -1718 & -2122 & -2178\\                 
\\
1edn  & 71\% & 53\% & 53\% & -113 & -107 & -97\\                 
\\
1epg  & 62\% & 51\% & 42\% & -313 & -382 & -230\\                 
\\

\end{tabular}}
\end{center}
\end{table}